# Quantum Correlations and Quantum Non-locality: a review and a few new ideas.

**Marco Genovese [1], Marco Gramegna [1]**

[1] INRIM - Istituto Nazionale di Ricerca Metrologica, Strada delle Cacce 91, 10135 Torino (Italy)

**Abstract:** In this paper we make an extensive description of quantum non-locality, one of the most intriguing and fascinating facets of quantum mechanics. After a general presentation of several studies on this subject, we consider if quantum non-locality, and the friction it carries with special relativity, can eventually find a "solution" by considering higher dimensional spaces.

## 1. Introduction

Bell discovery [1] that Local Hidden Variable Theories (LHVT) cannot reproduce all the results of quantum mechanics when dealing with entangled states, demonstrated the peculiarity of correlations among quantum states introducing the concept of quantum non-locality (NL).

Since then the experimental studies of Bell inequalities [2,13–15,17], recently culminated in conclusive tests [18], and the developing of quantum technologies [19] prompted a growing interest in this effect, which remains one of the most significant problems in understanding quantum mechanics.

The investigation on Bell NL properties has been developed and generalized in many directions, and in the present paper we aim to recall some of these achieved developments. To pursue this purpose, the first crucial point to be considered is about the definition of non-locality [2,26–30,53]. Looking at Bell theorem, it is possible to understand how **locality** entails with the fact that if two different observers, henceforth called Alice and Bob, make two independent measures (for example, A and B) on a shared physical system and achieve respectively the results $a$, $b$, the resulting joint probability of their measurement can be expressed in the following factored form

$$P(a, b|A, B, x) = P(a|A, x)P(b|B, x) \quad (1)$$

in which $x$ plays the role of a so called "hidden variable", endowed with its own probability density function $r(x)$, and acting an influence on the outcome in a deterministic or probabilistic way. This factorisation is dubbed Bell non-locality.

When this assumption is done, one can obtain different forms of Bell inequalities, as the Clauser-Horne one:

$$CH = P(a, b) - P(a, b') + P(a', b) + P(a', b') - P(a') - P(b) \leq 0 \quad (2)$$

that are verified for every LHVT, but violated in Quantum Mechanics. Therefore, this demonstrates that there exist quantum states whose outcome probabilities are inconsistent with locality.

Thus the non-locality property of entangled states leading to a violation of Bell inequality is violating the former rule (1). Very interestingly, after being a source of debate in the Foundations of quantum mechanics and epistemology, more recently Bell non-locality has then emerged as a fundamental resource for quantum technologies [71]. This further prompted several studies on quantum non-locality and a search for its understanding, and motivates reviewing some of the last studies in the following.

A further form of quantum non-locality was then provided by Mermin [4], extending GHZ [5,6] analysis. In this case one directly shows that there exist quantum states whose outcome possibilities are inconsistent with locality. In little more detail, by considering the GHZ state

$$\cos(\alpha)|000\rangle + \sin(\alpha)|111\rangle \qquad (3)$$

he demonstrated that no LHVT can reproduce the QM result for measurements $X_1 \otimes X_2 \otimes X_3$, $Y_1 \otimes Y_2 \otimes X_3$, $Y_1 \otimes X_2 \otimes Y_3$ and $X_1 \otimes Y_2 \otimes Y_3$. Interestingly, Mermin non-locality has also been recently included in categorical formulation of quantum mechanics [7]. Always in categorical formulation of quantum mechanics it has been also demonstrated [8] that strong contextuality (exemplified by Mermin/GHZ non-locality) implies possibilistic non locality (exemplified by Hardy non-locality [9]) that implies probabilistic non-locality (exemplified by Bell's inequalities).

It is interesting at this point to put in evidence that, aside important contributions on generalizations to non-deterministic HV theories, Stapp coined also an inspired and elegant definition of NL: "non-locality property of quantum mechanics is the logical need for information about a choice of experiment freely made in one region to be present in a second region that is space-like separated from the first" [72].

The hidden variable $x$ value is usually viewed as a "property" that the two measured physical systems posses because of their generation from a certain common source (for example for an entangled photon pairs emerging from a Spontaneous Parametric Down-Conversion crystal). If the two measurements are space-like separated events their outcomes exhibit a local dependence on the experimental settings and on $x$ itself.

Given these conditions, following Ref. [30] it is possible to demonstrate that these models fail when an experiment is designed to implement two distinct and independent sources generating entangled pairs, therefore entailing two independent sets $x_1$, $x_2$ of HVs, and the visibility overcomes the threshold of 50% in entanglement swapping [31] (it is worth to notice that in the case of experiments verifying Bell inequalities the required visibility is 71%).

More in general, $x$ can be explained as the state describing the two systems [26], or even the whole universe. In this case, referring to Eq. (1) the requirements on $A$, $B$ measurements imply that the dependence does not rely on $x$ only, i.e. $x$ can be a non-local *beable* with the only condition that the local inputs $A$, $B$ together with the global state $x$ determine the outcome probabilities

Moreover, in Ref. [21] it is argued that whether a deterministic model is considered, or epistemic probabilities are taken into account, problems arise related to independence of outcomes from reference frame.

Bell non-locality (1) makes reference to hidden variables $x$ ("ontological" version); in [22] operational definitions, i.e. signal locality (absence of signalling), $P(a|A, B) = P(a|A)$ for a certain preparation, and predicability (one can predict the outcomes of all possible measurements that be performed on a system), $P(a, b|A; B) \in 0, 1$ for a certain preparation, were shown to be sufficient for demonstrating Bell inequalities. Finally, it was demonstrated that it exists a complementarity for the resources needed for violating Bell inequalities. This can be written as $S + 2I > C$ both for ontic in [23] and operational cases [24,25], where $I$ and $S$ are randomness and signalling resources respectively $C$ being the average communication in bits: thus, non-local correlations (C>0) in presence of no signalling (S=0) requires upredictability (I > 0).

## 2. Various questions on quantum non-locality

Quantum nonlocality and entanglement exhibit a strict and deep interrelation, being at the same time two well distinct concepts, and the understanding of the differences in a quantitative terms deserves attention in order to clarify the balance between these two inequivalent resources. As a starting point, a first interesting aspect to be examined in detail in this chapter concerns whether QM violates relativistic locality, and on the consequence a superluminal influence between entangled entities can be a suitable explanation for quantum NL. Pursuing this purpose, a well-defined and sound model of this type was proposed in [13], introducing a preferred frame where faster-than-light signals propagate isotropically with unknown velocity $vt = \beta c$. It is important to notice that a preferred frame is not in contrast to special relativity and that a universal preferred frame has been already observed, being the cosmic microwave background. Ref. [39] posed a lower limit of more than ten thousand time the speed of light in the vacuum [39]. This limit has been recently enhanced [35] to $\beta > 5 \cdot 10^6$.

Moreover, the analysis reported in [40] shows that this hidden influence, although propagating to a subluminal speed in a preferred reference frame, does it allow always for superluminal communications: this peculiarity refers to 4-partite systems, but it is not valid for bipartite states. On the other side, this is not true for all the other models (including de Broglie-Bohm and GRW [2] for which the propagation in space of the correlations is not continuous. Moving to another aspect of the relation between entanglement and NL, we firstly mention an analytic demonstration that quantum nonlocality does not imply distillability of quantum correlations has been reported in Ref. [38] by considering a 3-qubit entangled state that is separable along any bipartition, but nevertheless violates a Bell inequality.

Moving to another interesting facet of the argument, the exact relation between entanglement and nonlocality is still poorly understood. In fact, in the reasoning of [41] it is demonstrated that a particular class of states can exhibit entanglement without violating Bell inequalities, in the sense that it is possible to obtain more nonlocality with less entanglement, not only at the qubit level [13] but also when there is no restriction on the size of the entanglement [42]; a situation experimentally investigated in [63]. An argument related to states that exhibit on one side a maximal non-locality without possessing a maximum degree of entanglement has been investigated in [44]. This study put in evidence three possible scenarios plausible with this situation: i) the case [13] in which states that are non-maximally entangled violate Bell inequalities when using single-particle detectors with non-ideal quantum efficiency below 82%; ii) for non-maximally entangled states, the Kullback-Leibler distance with the closest local distribution [45] is larger; iii) the simulation of entanglement with non-local resources is more favorable when maximally entangled states are considered rather then the non-maximally ones.

A further peculiarity lies in the evidence, discovered in [46], that the totality of non-fully-separable states with an arbitrary dimension in the Hilbert space and number of parties, violate a Bell inequality if mixed with another state that taken alone could not act a violation of the same Bell inequality.

From another study it appeared evident that an emerging quantum NL [47,48] is always unveiled by mixing a $\sigma_{AB}$ state not violating Bell inequalities with any entangled state $\rho_{AB}$, a peculiar trait still to be seized in detail for potential exploitation in quantum communication protocols. In [49] an interesting comparison is analyzed between the resource theories of NL (local operations and shared randomness, LOSR paradigm) and entanglement (local operations and classical communication, LOCC paradigm), and it is shown how a class of nonlocal games can witness quantum entanglement, however weak, and reveal nonlocality in any entangled quantum state. A study related to the quantification of nonclassicality and focused on the global impact of local unitary evolutions [50] puts in evidence that only for those composite quantum systems exhibiting quantum correlations (evaluated in the context of quantum-versus-classical paradigm [51]) the global state can undergo a transformation operated by means of any nontrivial unitary evolution; this property holds also when states non violating the Bell inequality are considered. More recently it was demonstrated [52] that all entangled states allow implementing a quantum channel, which cannot be reproduced classically, in particular teleportation according to a specific "teleportation witnesss". A relevant way to quantify nonlocality [54] takes into account the minimal amount of information necessary to preserve quantum correlation in a communication exchange between two parties, which in the particular case of a 2-qubit maximally entangled system undergoing projective measurements this quantification is accounted to be exactly 1 bit (on the contrary the situation can be different when the system is non-maximal entangled, a matter still under debate). On a different perspective is founded the argument detailed in [37], considering general hidden variable models that can be expressed in terms of both local and nonlocal parts, and demonstrating the existence of experimentally verifiable quantum correlations that are incompatible with any hidden variable model having a nontrivial local part. Following this reasoning, it is then conceivable that if the non-local HVs are not considered, measurement results rely exclusively on local HV parameters and experimental settings. In this sense, this is equivalent to saying that quantum mechanics cannot be mimicked by HV theories provided with a local component.

The following reported approach started by considering that in a typical Bell experiment, correlations are observed between the measurement results of Alice and Bob and averages are evaluated over measurements of many pairs of particles, but when nonlocality appears from statistics, this does not directly means that all individual pairs behave nonlocally. This inspired the reasoning [56] for which a set of the pairs behaves locally while another fraction in a non-local way. In other terms, it was argued that the relative quantum distribution probability $P_Q(a, b|A, B, \rho)$ for a given quantum state $\rho$ can be expressed in terms of local ($P_L$) and non-local ($P_{NL}$) parts, where respectively the $P_L$ satisfies Bell inequalities and the other $P_{NL}$ does not depend on outcomes or experimental settings:

$$P_Q = p_L(\rho) \cdot P_L + [1 - p_L(\rho)] \cdot P_{NL} \qquad (4)$$

A tricky problem resides in determining the weight of the local part (that represents in some sense the measure of locality for $P_Q$) and in finding a suitable way to minimize it. A step in this direction was the demonstration that the local part is identically null for maximally entangled 2-qubit systems, returning a complete non-local behaviour, while an investigation analyzing non-maximally pure 2-qubit and 3-qubit states can be found in [57]. Instead, focusing on the non-local component of Eq. (4) and making use of nonlocal resources uniquely, with no-signaling constraint, it is possible to simulate correlations for all pure 2-qubit entangled systems without communication [76], therefore exploiting no-signaling resources exclusively. A similar decomposition approach for contextuality has been validated in [58]. Taking advantage of the general way to express any q-qubit pure state in the form $|\Psi(\theta)\rangle = \cos(\theta)|00\rangle + \sin(\theta)|11\rangle$, Ref. [59] shows that $P_L = \cos(\theta)$ represents the maximal value
when for non maximally entangled states are considered.

The decomposition in Eq. (4) has been then used for an analysis in terms of non-locality distillation [77]. In this case, when a number of copies of a nonlocal system is defined, two relevant parameters can be identified: i) the non local cost and (ii) the distillable nonlocality. This study put in evidence the existence of nonlocal boxes showing what they define as bound nonlocality, i.e. whose distillable nonlocality is strictly smaller than their nonlocal cost, analyzing also whether the nonlocality contained in such boxes can be activated.

In analogy with this, in [78] Bell Inequalities are derived exploiting the known link between the Kochen-Specker and Bell theorems, and discovered other measurement bounds allowing a higher degree of nonlocal correlations than those reported for cases reported in.

Several further studies addressed measuring non-locality, for instance the study of interconversions among several copies of nonlocal resources was afforded in [79–81], the distillation of several copies of a nonlocal behaviour into a more nonlocal in [82], the operations under which nonlocality cannot increase used to find a proper definition of genuinely multipartite nonlocality [83].

A different approach, that overturns in some sense the perspective on how non-locality can be conceived, is built on reversing the logical order of Quantum Mechanics [60], by making nonlocality an axiom and indeterminism a theorem (QM relies on the opposite). In other words, it is argued that QM is not the most nonlocal theory (concerning with NL correlations) consistent with relativistic causality. This perspective gave the idea to account for nonlocal "superquantum" (known also as PR-box, by Popescu and Rohrlic, or NLB: non local box) correlations that interrelate without superluminal signalling the outputs of two parties with their inputs A,B in this way:

$$\sum_b P(a, b|A, B) = P(a|A) \quad \text{and} \quad \sum_a P(a, b|A, B) = P(b|B). \tag{5}$$

In Ref. [70] the properties of such kind of set of probabilities are detailed in terms of communication complexity.

For exemplifying PR box one can consider the situation when one has the relation between outputs and inputs

$$a \otimes b = A \cdot B \tag{6}$$

where $\otimes$ stands for addition modulus 2. Furthermore, Alice's and Bob's marginal distributions are completely random and relativity causality is satisfied (namely Alice and Bob cannot communicate directly by exploiting PR boxes.)

It is worth to mention that these nonlocal superquantum correlations generate a stronger violation ($S = 4$) in CHSH inequality with respect to any quantum correlations, remembering that Tsirelson's bound [61] gives $S = 2\sqrt{2}$ (Quantum Mechanics) and $S = 2$ in classical theory, while eventually a stronger violation [62,63] can be achieved with a larger number of measurements (chained inequalities) [64].

Substantially, quantum correlations do not violate locality in a stronger way as allowed by causality (at least for two dimensional systems, for higher dimensions the situation changes [65], using chained inequalities mentioned above). This raises the question if there is some physical principle limiting the correlations to QM ones. A partially negative answer was provided in [66], where it was shown as that quantum theory is, at the level of correlations among systems, not as special as expected identifying a set of non-signalling correlations (defined in terms of efficiently solvable semi-definite programmes) whose set is strictly larger than the quantum correlations one, but satisfies all-but-one (Information Causality, not yet proved) of the device-independent principles proposed for quantum correlations (Non-trivial Communication Complexity, No Advantage for Nonlocal Computation, Macroscopic Locality and Local Orthogonality). In a first approach, ref.[67], the problem was faced by a communication complexity point of view [70]. Taking into consideration the case of two parties dealing with a Boolean function $F(a; b)$, and for which each of the parts has knowledge of only one input (a or b), the interesting problem refers to the minimization in communication resources in order that a part can compute and solve the function. In general, the case in which a bit only is necessary results a trivial matter. For an analysis of nontrivial cases, it can be mentioned that [67] reported the comparison between the bounds for success probability, estimated to be: i) 90.8% in the case of PR-box operating in probabilistic way; and ii) 85.4% in the case Tsirelson case (considering that the classical level is fixed at 75%). The consequent partition of communication complexity problems in hard- and easy-solving classes can be understood in terms of partial characterization of non-local correlations by measurements on entangled systems. It is of interest the experimental verification [68] reporting that post-selected measurement results can approach the PR bound, while is well known that in Quantum Mechanics the limit is fixed by Tsirelson bound.

A study [75] focused on general properties relevant to the whole nonsignaling theories (predicting Bell inequalities violation) discerns among the following phenomena, that are a consequence of the no-signaling principle and nonlocality: intrinsic randomness, monogamy of correlations (in the sense that these correlations cannot be shared among a indefinite number of parties, and for this reason different with respect to classical analogue), impossibility of perfect cloning, uncertainty due to the incompatibility of two observables, privacy of correlations (if two honest parties know to share correlations with some degree of monogamy, they can estimate and possibly bound their correlations with a third untrusted party), bounds in the shareability of some states . Moreover, it was discovered that the properties of nonlocal, no arbitrarily shareable and positive secrecy contents result to be equivalent, and this turns out to be valid for any distribution.

Bell inequality violation implies that: (i) the considered quantum state must be entangled, (ii) the sets of local quantum measurements performed by one or the other party must be incompatible.

Can this link be reverted? As mentioned Werner demonstrated that entanglement does not imply violation of Bell inequalities.

One of the most recent line of investigation related with quantum non-locality is the connection with the absence of uncertainty free joint measurements of observable eventually used in Bell inequalities tests. In [85] it was shown that, for dichotomic POVM, every pair of incompatible quantum observables enables the violation of a Bell inequality and therefore must remain incompatible within any other no-signaling theory, in [86,87], further elaborated in [88,89], that a set of measurements is not jointly measurable if and only if it can be used for demonstrating EPR steering , in [90] that the result of [85] does not hold in general, as the joint measurability problem cannot be reduced to a pair of POVMs only (result extended in [91]). In [84] was established that the degree of nonlocality of any theory is determined by two factors: the strength of the uncertainty principle and the strength of steering, i.e. the possibility of explaining the Tsirelsons bound by exploiting the uncertainty limit of conditional quantum states deriving from the measurement of one system of a bipartite entangeld state. In particular it was examined the game where Alice and Bob win when their answer satisfies relation 6 between outputs and inputs. Then they considered the concept of **steerability**, i.e. what states Alice can prepare on Bob's system remotely. When they share a state $\rho_{AB}$, the reduced density matrix pertaining Bob's system is

$$\rho_B = \Sigma_a P(a|s) \, \rho_{s,a} \tag{7}$$

Schrödinger noticed that for all $s$ there exist a measurement on Alice's system that allows steering the state $\rho_{s,a}$ with probability $p(a|s)$ on Bob's site. This steering procedure does not allow to violate no-signalling because for each of Alice's' settings, Bob'state is the same when averaging over Alice's measurements outcomes. The tight bound for the game success is:

$$P^{game} = \Sigma_s p(s) \, \Sigma_a P(a|s) \, \zeta_{s,a} \tag{8}$$

where $\zeta_{s,a}$ is a proper of uncertainty relation in entropic form [84] (one cannot obtain a measurement outcome with certainty for all the measurements simultaneously whenever $\zeta_{s,a} < 1$).

On the other hand, Ref.[93] (see also [97]) proved a link between Tsirelsons bound and measurement incompatibility in two qubits Hilbert space (i.e. with the optimal degree of unsharpness).

By analysing covariance matrix Ref.[94] demonstrated the inequality

$$|B| \leq 22 + Min[S(a), S(b)] \tag{9}$$

where $B$ is the Bell-CHSH parameter and $S(a)$, $S(b)$ are Tsallis entropies concerning local measurements of Alice and Bob. The use of weak values [95] was suggested for experimentally testing these inequality. Ref. [96] characterised qiuantum correlations in terms of generalized uncertainty relations as well as specific notion of locality (local uncertainty relations cannot be affected by spacelike separated events).

Finally, Ref. [93] carried on this path demonstrating a bound on non-local correlations (stronger than Tsirelson one) in terms of local measurements of two independent joint measurements on two quantum correlated systems, in particular the requirement of positivity for experimental probabilities measured in a combination of two local joint measurements realized on two separate quantum systems leads to a bound on non-local correlation between these.

The fact that steering occurs for mixed entangled states that are Bell-local has been experimentally demonstrated [101] by observing that a violation occurs when specific inequalities are considered (see also [103]), which extends previous works [102] relying on inferred variances of complementary observables [105] (also when Transition-Edge Detectors, that are intrinsically Single-particle number resolving, are experimentally implemented to avoid detection loophole [104]).

However, the inequalities considered in [105] are not violated by some particular entangled states: Ref. [106] describes and experimentally test a new criterion founded on entropy functions, violated by more states.

In [110] it was investigated whether the phenomenon of EPR steering can also be generalized beyond quantum theory demonstrating that, whilst post-quantum steering does not exist in the bipartite case, it is present in the case of three observers and it is fundamentally different from postquantum nonlocality.

A discussion on the connections between non-locality and complementarity can be found in [99].

The key point consists of the possibility to attribute to the concept of joint measurability of two observables (A,B) different operational meanings : i) A and B values can be obtained by the measurement of a third observable C (in the sense that the probability distribution of A and B outcomes are included as marginals in the distribution of C results) ii) when the sequence ABA is measured, the two measurements of the same observable A return identical outcomes iii) if the measurement sequence AB is considered, the same probability distribution over B outcomes is obtained as a direct B measurement.

In the context of Quantum Mechanics there is a total equivalence among all these statements, nevertheless Ref.[99] describes that for any no-signalling theory not showing complementarity (as specified at item iii) it is allowed a local realistic description, but this is not valid with respect to properties i and ii (the demonstration took advantage of NLB correlations models)

We also mention the research that analyzes the comparison among the structures of entanglement correlations with respect to the non-local ones, putting in evidence significant differences. It is there [73] speculated the concept of "unit of non-locality" as a possible explanation for a PR-box, and proven that this is not applicable to describe correlations when cluster states are considered. In [74] an investigation on other aspects of PR-Box is extended. Moreovew, the paper [111] analyzes the possibility to build a covariant deterministic nonlocal HV extensions of quantum theory, and demonstrates how it is possible to reduce any covariant nonlocal variables to Bell local variables whose existence is known to be incompatible with well-tested quantum predictions.

Another fundamental problem considered in Ref. [112] is related to whether the nonlocality of a quantum state can be superactivated, in other words if a state $\rho \otimes \rho$ can be considered non local if $\rho$ is local. This aspect, that refers to a sort of additivity property of NL, has been analyzed by considering unbounded [113] Bell inequalities, returning the conclusion that if a number of local entangled states are combined by direct product, it is possible to get a non-local global state, falsifying in this way the forementioned additivity. In conclusion, as an extension of this study it was considered the possibility for any kind of local entangled state to generate NL superactivation, and demonstrated [114] that all quantum states useful for teleportation represent nonlocal resources.

**3. Non-locality in higher dimensional spaces**

After having discussed in some detail quantum non-locality, it emerges the question of its compatibility with special relativity. Even if one can rigorously demonstrate that it can not lead to any superluminal transmission (signalling) [10], peaceful coexistence between special relativity and QM would require more, i.e. it would be necessary understanding a coherent description among different observers [115]. For instance, how to reconciliate two observers that observe a different temporal order in the collapse of two entangled particles? The answers to this question span from being an unsolved problem [116], requiring a preferred foliation to relativistic space-time, to being only apparent, since accounts of entangled systems undergoing collapse yielded by different reference frames can be considered as no more than differing accounts of the same process and events: there is a form of holism associated to QM description of composite systems, the factorizable state after collapse on a certain hypersurface is merely one way of slicing together local parts, entangled states are superposition of such splicings.

Be that as it may, rebus sic stantis, an undoubted difficulty in understanding the collapse in composite systems persists. Our suggestion is that one should explore the possibility that this problem can be solved in higher dimensional spaces. Let us consider a $n > 4$ space-time with $n − 4$ more spatial dimensions and let us suppose that, while the usual fields only "leave" in 3 + 1 dimensions, the collapse is mediated by "some field" propagating also in the other dimensions. Since every n-dimensional space can be "contracted" around a single point of a higher dimensional space (for example a plane can be wrapped as tied as one wishes around a single point in 3 dimensional space), two far points in the "ordinary" space can be as close as wanted considering extra-dimensions. Incidentally, this "contraction" is an isometry and leaves invariant gaussian curvature.

Thus, if this higher dimensions are extremely small, every event of the 3+1 dimensional space could be connected with any another event in an "extremely short" effective distance through the compactified dimensions. Thus, the collapse of one of two remote entangled particles could cause the collapse of the other with the propagation of a subluminal signal through the compactified dimensions. Quantum non-locality would reduce to the fact that only wave function collapse would be affected by these extra dimensions.

This kind of phenomenon could either be included in a more general theory going beyond quantum mechanics (for example some Planck scale theory already predicts compactified dimensions) or find a use in building relativistic collapse models, for example a simple generalization of Tumulka's one [118] should be able to incorporate it.

For instance, one could think to some general non-unitary evolution operator

$$U(t) = Exp\{-i\int d^4x\, d^n y O_I(x,y,t)\} \qquad (10)$$

where $O_I(x, y, t) = H_I(x, t) + C_I(x, y, t)$, $H_I(x, t)$ being the usual Hermitian Hamiltonian density of quantum fields leaving in a 3+1 dimensional space (x), while $C_I(x, y, t)$ is not Hermitian and eventually induces the collapse. When one traces over the degrees of freedom corresponding to the not Hermitian component, $C_I(x, y, t)$, the usual unitary evolution is eventually restored, as in 't Hooft finite degrees of freedom model [119]. Finally, it would be worth investigating if this extra component could contribute to solve other issues of modern physics, as dark matter and dark energy.

## 4. Acknowledgements

This research was supported by grant number (FQXi FFF Grant number FQXi-RFP-1812) from the Foundational Questions Institute and Fetzer Franklin Fund, a donor advised fund of Silicon Valley Community Foundation. This work has received funding from PATHOS EU H2020 FET-OPEN grant no. 828946, and from the European Unions Horizon 2020 and the EMPIR Participating States in the context of the project EMPIR-17FUN01 'BeCOMe'.## 5. Bibliography

1. J.S. Bell, Physics 1 (1965) 195.
2. M. Genovese, *Phys. Rep.* 413/6, 319-398 (2005); Adv. Sci. Lett. 3, 249258 (2010) and references therein.
3. D.Ahn et al., arXiv 1207.6062; Y.Agharonov et al, arXiv 1206.6224; S.Lloyd et al., Phys.Rev.Lett.106 82011) 040403; O.Oreshkov et al.Nat.Comm. (2012) 10.1038; W.K. Wootters, Int.Jour. Th. Phys 23 (84) 701;
   M.Genovese, Advanced Science Letters 2, (2009) 303;
4. N.D.Mermin, Am. J. Phys. 58 (1990) 731.
5. D.M.Greenberger, M.A. Horne and A. Zeilinger, in "Bell's Theorem, Quantum Theory and Conceptions of the Universe", ed. M. Kafatos, (Kluwer, Dordrecht, 1989), 73.
6. D.M. Greenberger, M.A. Horne, A. Shimony, and A. Zeilinger, American Journal of Physics 58, 1131 (1990).

7. B.Coecke et al., arcXiv1203.4988
8. New J.of Phys 13 (2011) 113036.
9. L.Hardy, PRL 71 (1993) 1665.
10. G. C. Ghirardi, A. Rimini and T. Weber, Lett. Nuov. Cim. 27 (1980) 293.
11. W. Tittel et al., PRA 59 (1999) 4150.
12. G. Weihs et al., PRL 81 (1998) 5039.
13. P. H. Eberhard, *PRA* **47**, R747, 2800-2811 (1993).
14. G. Brida et al., *PLA* **268**, 12-16 (2000).
15. M.A. Rowe et al., Nature 409 (2001) 791.
16. E. Schrödinger, Proc. of Camb. Phyl.
17. M. Giustina et al., Nature 497, 227-230 (2013).
18. B. Hensen et al., Nature (London) 526, 682 (2015). L. K. Shalm et al., Phys. Rev. Lett. 115, 250402 (2015). M. Giustina et al., Phys. Rev. Lett. 115, 250401 (2015).
19. " Quantum Information,Computation and Cryptography", (Springer, Berlin, 2010), Editors: F.Bennati et al.; M.Genovese, Journal of Optics, 18 (2016) 073002.
20. H.Nikolic, International Journal of Quantum Information 15, 1740001 (2017).
21. N.Gisin, arXiv1012.2536
22. E.G. Cavalvanti and H.M.Wiseman, Found. of Phys. 42, 1329 (2012).
23. M.Hall,PRA 82, 062117 (2010).
24. S.Aravinda and R. Srikanth, J.Phys A 49, 205302 (2016) .
25. S.Aravinda and R. Srikanth, , arXiv 1811.12409.
26. N.Gisin arXiv1012.2536
27. A. J. Leggett, Found. of Phys. 33, 1469 (2003).
28. S.Gröblacher et al., Naure 446, 871 (2007).
29. C.Branciard, et al., Nature Phys. 4, 681 (2008).
30. M.Zukowsky et al., PRL 71, 4287 (1993).
31. C.Branciard et al., PRL 104, 170401 (2010).
32. P. H. Eberhard, in Quantum Theory and Pictures of Reality: Foundations, Interpretations, and New Aspects, edited by W. Schommers (Springer, Berlin, 1989), p. 169.
33. D. Bohm and B. J. Hiley, The Undivided Universe: An Ontological Interpretation of Quantum Mechanics (Routledge, London, 1993).
34. B. Cocciaro, J. Phys.: Conf. Ser. 626, 012054 (2015).
35. B. Cocciaro et al., Phys. Rev. A 97, 052124 (2018).
36. J.Barret and N.Gisin, PRL 106 (2011) 100406..
37. R.Colbech and R.Renner, PRL 91, 050403 (2008).
38. T.Vertesi and N. Brunner, PRL 108 (2012) 030403.
39. D.Salart et al., Nature 454 (2008) 861.
40. J. Bancal et al. Nat. Phys. 8 (2012) 867.
41. R.F. Werner, PRA 80, 4277 (1989).
42. Y.-C. Liang, T. Vertesi, and N. Brunner, Phys. Rev. A 83, 022108 (2011); M. Junge and C. Palazuelos, Commun. Math. Phys. 306, 695 (2011); T. Vidick and S. Wehner,Phys. Rev. A 83, 052310 (2011).
43. B.G.Christensen et al., Phys. Rev. X 041052 (2015).
44. A.A.Methot and V.Scarani, Quant. Inf. Comput 7 (2007) 157.
45. A.Acin et al, PRL 95 (2005) 210402.
46. Y. Liang et al., arXiv:1210.0548.
47. L. Masanes, Y.-C. Liang, and A. C. Doherty, Phys. Rev. Lett. 100, 090403 (2008).
48. M. Navascues and T. Vertesi, Phys. Rev. Lett. 106, 060403 (2011).
49. 4.F. Buscemi, Phys. Rev. Lett. 108, 200401 (2012).
50. S. M. Giampaolo et al., Phys. Rev. A 87, 012313 (2013).

51. B. Dakic, V. Vedral, and C. Brukner, Phys. Rev. Lett. 105,190502 (2010).
52. D. Cavalcanti, P.Skrzypczyk, and I. Supic, Phys. Rev. Lett. 119, 110501 (2017).
53. H.M. Wiseman et al., PRL 98, 140402 (2009).
54. G.Brassard et al., PRL 83, 1874 (1999); M.Steiner,PLA 27 0, 239 (2000).
55. B.F.Toner and D.Bacon, Phys. Rev. Lett. 91, 187904 (2003).
56. A.Elitzur, S.Popescu and D.Rohrlich, PLA 162, 25 (1992).
57. V.Scarani, PRA 77, 042112 (2008); Branciard C, Gisin N, Scarani V.,PRA 81, 022103 (2010).
58. E. Anselem et al.,arXiv:1111.3743v2
59. S.Portmann,C.Branciard,N. Gisin, PRA 86, 012104 (2012).
60. S.Popescu and D.Rhorlich, Found. Phys. 24, 379 (1994).
61. B.S. Tsirelson, Lett. Math. phys. 4, 93 (1980).
62. S.Wehner, PRA 73, 022110 (2006).
63. B.Christensen et al., PRX 5, 041052 (2015)
64. S. Braunstein and C. Caves, ANNALS OF PHYSICS 202, 22 (1990)
65. A.Cabello, PLA 337 (2012) 64.
66. M.Navascues et al., Nat. Comm. 6 6288 (2015).
67. G.Brassard et al., PRL 96 ,250401 (2006).
68. N.Gisin, PLA 210, 151 (1996);S.Marcivitch et al., PRA 75, 022102 (2007); Y.Chen et al., PRL 97, 170408 (2006); A.Cabello, PRL 88, 060403 (2002).
69. D.S.Tasca et al., PRA 80, 030101 (2009).
70. H.Buhrman et al., Rev. Mod. Phys. 82, 665?698 (2010)
71. J.Barrett et al., Phys. Rev. A 71, 022101 (2005).
72. H.P. Stapp, Am.Journ. of Phys 72 (2004) 30; Found. of Phys. 10701-012-9632.
73. J.Barrett and S.Pironio, PRL 95, 140401 (2005).
74. A.Short et al., Phys. Rev. A 73, 012101 (2006); J.Barrett, Phys. Rev. A 75, 032304 (2007); I.Pitowsky, Phys. Rev. A 77, 062109 (2008); N.Linden et al., Phys. Rev. Lett. 99, 180502 (2007);
75. L.Masanes, Phys. Rev. A 73, 012112.
76. N.Brunner, Phys. Rev. A 78, 052111 (2008).
77. N.Brunner et al., PRL 106, 020402 (2011).
78. L.Aolita et al., PRA 85 (2012) 032107.
79. N. S. Jones and L. Masanes, Phys. Rev. A 72, 052312 (2005)
80. J. Barrett and S. Pironio, Phys. Rev. Lett. 95, 140401 (2005)
81. Forster M and Wolf S 2011 Phys. Rev. A 84 042112
82. Forster M, Winkler S and Wolf S 2009 Phys. Rev. Lett. 102 120401; Brunner N and Skrzypczyk P 2009 Phys. Rev. Lett. 102 160403; Brunner N, Cavalcanti D, Salles A and Skrzypczyk P 2011 Phys. Rev. Lett. 106 020402 83.  Gallego R., Wurflinger L E, Acin A and Navascues M 2012 Phys. Rev. Lett. 109 070401
84. J.Oppenheim and S.Wehner, Science 330, 1072 (2010).
85. Michael M. Wolf, David Perez-Garcia, and Carlos Fernandez Phys. Rev. Lett. 103, 230402
86. M. T. Quintino, T. Vertesi, and N. Brunner, Phys. Rev. Lett. 113, 160402
87. R. Uola, T. Moroder, and O. Ghne, Phys. Rev. Lett. 113, 160403
88. Neil Stevens and Paul Busch, Phys. Rev. A 89, 022123
89. H. S. Karthik, A. R. Usha Devi, and A. K. Rajagopal, Phys. Rev. A 91, 012115
90. M. T. Quintino, J. Bowles, F. Hirsch, and N. Brunner, Phys. Rev. A 93, 052115
91. F. Hirsch, M. T. Quintino, and N. Brunner, Phys. Rev. A 97, 012129
92. J. Oppenheim and S. Wehner, Science 330 (2010) 1072
93. M.Banik et al., PRA A 87, 052125 (2013)
94. A. Carmi and E. Cohen, entropy 20 (2018) 500.
95. A. Aharonov et al., PRL 60 (1988) 1351; F.Piacentini, A.Avella, M.Gramegna, R.Lussana, F.Villa, A.Tosi,

G.Brida, I.Degiovanni and M.Genovese, Scientific Reports 8 (2018) 6959; F. Piacentini, A. Avella, M.P. Levi, M. Gramegna, G. Brida, I.P. Degiovanni, E. Cohen, R. Lussana, F. Villa, A. Tosi, F. Zappa, and M. Genovese; Phys. Rev. Lett. 117 (2016) 170402.

96. A. Carmi and E. Cohen, arXiv 1806.03607
97. M. Bush et al., EPL, 103 (2013) 10002 98.    H. Hofmann, arXiv:1809.04725
99.  T. Fritz, PRA 85 (2012) 022102.
100. Cary P., Behav. Scienc. & the Law 25, 165 (2007).
101. D. J. Saunders, S. J. Jones, H. M. Wiseman and G. J. Pryde, Nature Physics, 6, 845 (2010)
102. Ou, Z. Y., Pereira, S. F., Kimble, H. J. Peng, K. C. Phys. Rev. Lett. 68, 3663 (1992); Bowen, W. P., Schnabel, R., Lam, P. K. and Ralph, T. C; Phys. Rev. Lett. 90, 043601 (2003); Hald, J., Sorensen, J. L., Schori, C. and Polzik, E. S. Phys. Rev. Lett. 83, 13191322 (1999); Howell, J. C., Bennink, R. S., Bentley, S. J. and Boyd, R. W. Phys. Rev. Lett. 92, 210403 (2004).
103. H.Meng et al., International Journal of Quantum Information 16, 1850034 (2018) 104. H. Smith et al., Nat. Comm. DOI:10.1038/ncomms1628.
105. M.D.Reid, PRA 40, 913 (1989); M.D.Reid et al., Rev.Mod.Phys. 81,1727 (2009).
106. S.P. Walborn et al., PRL 106, 130402 (2011).
107. A. Einstein, B. Podolsky and N. Rosen, Phys. Rev. 47, 77 (1935).
108. E. Schrödinger, Die Naturwissenschaften 23, 807 (1935).
109. J.S. Bell, Rev.Mod. Phys. 38, 477 (1966).
110. A.Sainz et al., PRL 115, 190403 (2015).
111. N.Gisin, PRA 83, 020102(2011).
112. C. Palazuelos, Phys. Rev. Lett. 109, 190401 (2012).
113. M. Junge, C. Palazuelos, D. Perez-Garcia, I. Villanueva, and M. M. Wolf, Phys. Rev. Lett. 104, 170405 (2010).
114. D. Cavalcanti, A. Acin, N. Brunner, and T. Vertesi, Phys. Rev. A 87, 042104 (2013).
115. Hyomony A, Natural Science and Mathaphysics, Cambredge Univ. Press. 1993.
116. T.Maudlin, Pace Time in the quantum world, In J.T. Cushing et al., Bohmian Mechanics and quanstum theory: an appraisal, 1996 Kluwer Ac. Pub.
117. W.C. Myrvold, Stud. in Hyst. and Phyl. of Mod. Phys 33 (2002) 435.
118. R. Tumulka, quant-ph 0508230, Proc. Roy. Soc.Lond. A462 82006) 1897. 119. G. 't Hooft, Class. Quant. Grav. 16 (1999) 3263.